\documentclass[prl,twocolumn,aps,showpacs,superscriptaddress]{revtex4}
\usepackage{graphics}
\usepackage{epsfig}
\usepackage{amsmath}
\usepackage{amssymb}
\usepackage{color}

\newcommand{\rom}[1]{\uppercase\expandafter{\romannumeral #1\relax}}

\begin{document}
\newcommand{\beq}{\begin{equation}}
\newcommand{\eeq}{\end{equation}}

\title{Origami building blocks:  generic and special 4-vertices}

\author{Scott Waitukaitis}
\affiliation{Huygens-Kamerlingh Onnes Lab, Leiden University, PObox 9504, 2300 RA Leiden, The Netherlands}\affiliation{FOM Institute AMOLF, Science Park 104, 1098 XG Amsterdam, The Netherlands}
\author{Martin van Hecke}
\affiliation{Huygens-Kamerlingh Onnes Lab, Leiden University, PObox 9504, 2300 RA Leiden, The Netherlands}\affiliation{FOM Institute AMOLF, Science Park 104, 1098 XG Amsterdam, The Netherlands}

\date{\today}

\begin{abstract}
Four rigid panels connected by hinges that meet at a point form a 4-vertex, the fundamental building block of origami metamaterials. Here we show how the geometry of 4-vertices, given by the sector angles of each plate, affects their folding behavior. For generic vertices, we distinguish three vertex types and two subtypes.  We establish relationships based on the relative sizes of the sector angles to determine which folds can fully close and the possible mountain-valley assignments.  Next, we consider what occurs when sector angles or sums thereof are set equal, which results in 16 special vertex types.  One of these, flat-foldable vertices, has been studied extensively, but we show that a wide variety of qualitatively different folding motions exist for the other 15 special and 3 generic types.  Our work establishes a straightforward set of rules for understanding the folding motion of both generic and special 4-vertices and serves as a roadmap for designing origami metamaterials.
\end{abstract}

\maketitle

In recent years physicists, mathematicians, artists and engineers have shown that origami provides a robust design platform in realms as diverse as robotics \cite{Felton:2014, Wood:2010jz}, medical devices \cite{Kuribayashi:2006}, battery optimization \cite{Zeming:2013} and mechanical metamaterials \cite{Schenk:2013kk, Wei:2013kn, Silverberg:2014, Cheng:2014, Waitukaitis:2015, Evans:2015, Yasuda:2015, Dias:2012, Silverberg:2015, Na:2015, Cheung:2014, Liua:2015, Lechenault:2014, Hanna:2014}.  In regard to the last category, Schenk and Guest \cite{Schenk:2013kk} illustrated how the Miura-ori fold tessellation, a pattern originally designed for compactifying space cargo \cite{Miura:1985tt}, is in fact an auxetic 2D metamaterial.  Concurrently, Wei et al.~confirmed this result and also connected the geometry of the Miura-ori to the physics by dressing the folds with torsional springs and revealing nonlinear rigidity and bending response.  Silverberg et al., again using Miura-ori, discovered that its stiffness is tunable with the introduction of ``pop-through'' defects \cite{Silverberg:2014}, while Evans et al.~recast tessellated origami into the language of conventional lattice mechanics \cite{Evans:2015}.
Using a more generalized pattern, we showed how mechanical frustration of origami can lead to multistable metamaterials \cite{Waitukaitis:2015}.

\begin{figure}[b!]
\includegraphics[]{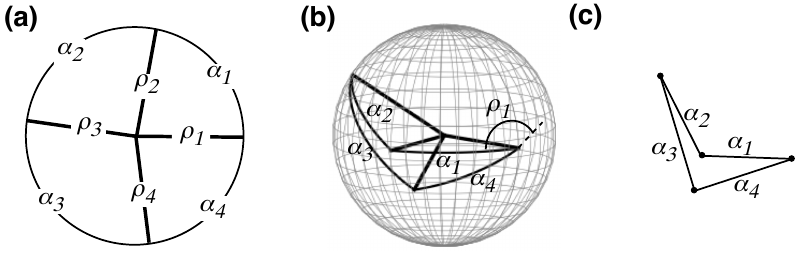}
\caption{Geometry of a 4-vertex.  (a) Illustration of 4-vertex geometry with sector angles $\alpha_i$ and folds $\rho_i$. (b) Illustration showing that the outer edges of the plates form great circles on the unit sphere, as well as an example of the fold angle $\rho_1$.  (c)  Schematic 2D drawing of the folded vertex from (b) as a 2D polygon.}
 \label{fig:vertex_geometry}
\end{figure}

The key ingredient in origami-based functionality is the intimate connection between geometry and folding motion.  In all of the aforementioned origami metamaterials, the mechanics is governed by one fundamental unit---the 4-vertex.  This is a structure where four rigid, wedge-shaped plates (with sector angles $\alpha_i$) surrounded by four folds meet at a point---see Fig.~\ref{fig:vertex_geometry}(a).  The 4-vertex is the simplest rigid-plated unit one can consider because it has just one continuous degree of freedom---vertices with fewer folds are fixed.  Historically, the focus has been on flat-foldability, i.e., the situation where all folds can close simultaneously. In 4-vertices, this is possible when the sum of alternating sector angles is equal ($\alpha_1+\alpha_3 = \alpha_2 + \alpha_4$) \cite{Kawasaki:1989, Justin:1989, Demaine:2007}.  Most studies thus far have focused on one particular pattern, the Miura-ori, which is based on a 4-vertex that, in addition to being flat-foldable, is also highly symmetric.
As we demonstrated, basing origami metamaterials on such a restricted class of 4-vertices is not necessary---a large space of 4-vertex geometry remains to be explored.  

Motivated by the connection between metamaterial functionality and geometry, we fully characterize the space of Euclidean 4-vertices, i.e., those whose sector angles add to $2\pi$.  We answer many fundamental but largely unaddressed questions:  How can 4-vertex geometries be categorized?  What are the possible mountain-valley arrangements?  Which folds can fully close?  How many ways can the vertex be folded? What roles do symmetries play?

Similar to what has been done for flat-foldable vertices, we show that much progress can be made by considering maximally folded states, and that the answers to these questions can be expressed via relationships between the sector angles.  First we focus on {\it generic} 4-vertices, i.e., those that have no special relationships between the sector angles, where we identify three distinct types.  We establish simple rules that determine the possible mountain-valley arrangements and the folds that can fully close for each type.  Moreover, we show that for each type there are two possible subtypes with qualitatively different folding behavior.

Next we move on to special 4-vertices, where we allow pairs of sector angles or sums of pairs to be equal.  Flat-foldable vertices arise naturally in this scheme, but we show that they are just one of 16 distinct special types.  These can be distinguished based on their codimensionality within the space of sector angles:  nine codimension-1 vertices, six codimension-2 vertices, and finally one codimension-3 vertex in which all sector angles are equal.  These special types partition the space of generic 4-vertices, dividing it into regions of differing generic types and subtypes.  Finally, we show that eight of these have fundamentally different folding branches than generic 4-vertices \cite{Waitukaitis:2015}.

As we discuss and analyze the different generic and special vertex types, we invite the reader to fold example 4-vertices from paper (conveniently available as cutouts in the supplemental material \cite{supplCutouts}) to illustrate how subtle differences in the flat geometry lead to different folding motions.

\begin{figure}[t!]
\includegraphics[]{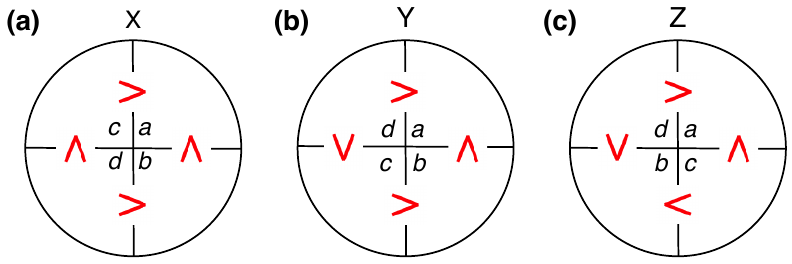}
\caption{Generic 4-vertices.  The three types of generic vertices, X, Y and Z, are determined by the arrangement of the ordered sector angles, $a<b<c<d$, around the vertex center.  In addition to the letters, the less than/greater than signs also indicate the relative sizes of adjacent sector angles.}
 \label{fig:genericTypes}
\end{figure}

\section{Generic 4-vertices}

As shown in Fig.~\ref{fig:vertex_geometry}(a), a 4-vertex consists of four rigid plates with sector angles $\alpha_i$ ($i = 1,2,3,4$, counted counterclockwise) connected by four folds (or hinges). A partially folded state of the vertex is captured by the values of the four fold angles $\rho_i$, defined as the deviation from in-plane alignment between neighboring plates [$\rho_i<0$ for ``valleys'', $>0$ for ``mountains'', and $=0$ for unfolded---see Fig.~\ref{fig:vertex_geometry}(b)]. We assume that $\sum\alpha_i=2\pi$ and that any one sector angle is smaller than the sum of the other three ($\alpha_j<\sum_{i\ne j}\alpha_i$---otherwise the vertex motion is trivial \cite{footnote1}). If we take the length of the folds shown in Fig.~\ref{fig:vertex_geometry}(a) as unity, the outer rims of the plates create great circles on the unit sphere [Fig.~\ref{fig:vertex_geometry}(b)].  Keeping the counterclockwise orientation of the vertex in mind, any folded state of a 4-vertex is thus equivalent to an oriented 4-polygon on the unit sphere \cite{Huffman:1976}.  For convenience, we will typically represent this with an oriented 4-polygon in the plane, as in Fig.~\ref{fig:vertex_geometry}(c).  Our analysis of folding motion will be primarily based on spherical triangle inequalities, which take the same form for the spherical and planar representations of Fig.~\ref{fig:vertex_geometry}(b) and (c).

Generic 4-vertices are those for which no special relationships exist between the sector angles.  Consequently there exists a well-defined ordering of the sector angles $a<b<c<d$ [for the example in Fig.~\ref{fig:vertex_geometry}(a), $\alpha_1=a<\alpha_4=b<\alpha_2=c<\alpha_3=d$].  The distinct types of generic 4-vertices correspond to different arrangements of these ordered angles around the vertex center, which we can represent schematically by placing the letters $a$ through $d$ in four quadrants of a circle, as in Fig.~\ref{fig:genericTypes}.  At first glance it appears there are $4!=24$ such arrangements, but this is reduced when one considers inherent symmetries.  We account for discrete rotational symmetry by putting the smallest angle $a$ in the upper right quadrant---this reduces from 24 to six.  Second, we account for the symmetry associated with flipping the vertex over by choosing the angle in the upper left quadrant to be larger than the one in the lower right quadrant.  This leaves just three types of generic vertices that cannot be mapped onto each other by reflection or rotations. We call these type X, Y and Z, as shown in Fig.~\ref{fig:genericTypes}.

\begin{figure}[t!]
\includegraphics[]{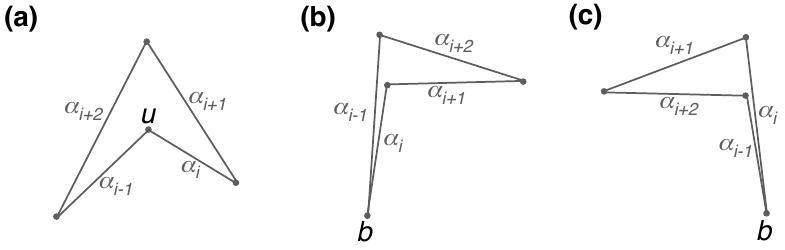}
\caption{Unique and binding folds.  (a) Schematic illustrating that if fold $i$ is a unique fold it must satisfy $\alpha_{i-1}+\alpha_{i}<\alpha_{i+1}+\alpha_{i+2}$.  If fold $i$ is a binding fold it must satisfy the spherical triangle inequalities with angles $\{ \alpha_{i-1}-\alpha_i, \alpha_{i+1}, \alpha_{i+2} \}$ (b) or $\{\alpha_{i}-\alpha_{i-1}, \alpha_{i+1}, \alpha_{i+2} \}$ (c). }
 \label{fig:uniqueBinding}
\end{figure}

\subsection{Generic folding behavior}

Given a particular generic 4-vertex, how can we expect it to fold?  In this section we focus on establishing the answer to this question. By introducing the notions of  the unique and binding folds, the unique and binding plates, and the dominant pair, the possible folding motions of generic 4-vertices can be expressed conveniently.  This leads to the conclusions that (1) generic 4-vertices have two branches of folding motion and (2) each generic type can be further categorized into two distinct subtypes.

\paragraph{Unique folds:}
As can be verified by folding the generic vertices included in the supplemental material \cite{supplCutouts}, some folds are capable of ``cupping'' into the others and having a fold angle with the opposite sign from the rest \cite{Huffman:1976, Waitukaitis:2015}.  [In Fig.~\ref{fig:vertex_geometry}(b), fold $\rho_2$ between plates $\alpha_1$ and $\alpha_2$ exhibits this cupping behavior.]  We call a fold capable of doing this (not necessarily doing this in a given configuration) a {\it unique fold}. As illustrated in the 2D schematic of Fig.~\ref{fig:uniqueBinding}(a), fold $i$ is a unique fold if the sum of the sector angles adjacent to it is smaller than the sum of the remaining two, i.e.,
\begin{equation}
\alpha_{i-1}+\alpha_{i}<\alpha_{i+1}+\alpha_{i+2}~.
\label{eq:unique1}
\end{equation}
In general, it follows from Eq.~(\ref{eq:unique1}) that if fold $i$ is a unique fold, then the opposing fold $i+2$ is not and vice versa. Moreover, this means that either fold $i+1$ or fold $i-1$ (but not both) is also a unique fold. Hence, a generic 4-vertex will always have two unique folds, labeled $u$ and $u'$, that straddle a common plate  \cite{footnote2}.  We call this the {\it unique plate} and designate it by the letter $U$.  The inequality in Eq.~\ref{eq:unique1} has a simple geometrical interpretation that allows one to quickly identify the unique folds by inspection (i.e.,~by asking ``which pair of neighboring plates is smallest?'').  For example, for the vertex shown in Fig.~\ref{fig:vertex_geometry}(a) folds 1 and 2 are the unique folds and plate 1 is the unique plate.

The existence of two unique folds has an important implication:  generic 4-vertices have two branches of folding motion that emerge from the flat state.  This is because two folds cannot have the opposite sign from the rest simultaneously.  For generic vertices, each branch of folding motion will have a corresponding unique fold.  As we will show later, this changes for certain special vertices.

\paragraph{Binding folds:}
Another observation readily made while folding a generic 4-vertex is that some folds are capable of fully closing to $\pm \pi$ while others are not.  We call a fold that is capable of fully closing a {\it binding fold}.  As shown in Figs.~\ref{fig:uniqueBinding}(b) and \ref{fig:uniqueBinding}(c), closing such a fold creates a spherical triangle.  If fold $i$ binds and $\alpha_i<\alpha_{i-1}$ then the side lengths of the spherical triangle are $\{\alpha_{i-1}-\alpha_i, \alpha_{i+1}, \alpha_{i+2}\}$.  Otherwise, if $\alpha_i>\alpha_{i-1}$ they are $\{\alpha_{i}-\alpha_{i-1}, \alpha_{i+1}, \alpha_{i+2}\}$.  In either case, the three sides must obey the three permutations of the spherical triangle inequality, which for generic vertices take the form of strict inequalities,
i.e.,
\begin{align}
\alpha_i& \!<\!\alpha_{i\!-\!1}:\begin{cases}
                                      \alpha_{i\!-\!1}\!-\!\alpha_{i}  \!<\! \alpha_{i\!+\!1}\!+\!\alpha_{i\!+\!2}  ~~~~~~~~~(i)\\
                                      \; \; \; \; \;  \; \;\alpha_{i\!+\!1}  \!<\! \alpha_{i\!+\!2}\!+\!\alpha_{i\!-\!1}\!-\!\alpha_i ~~~(ii)\\
                                      \; \; \; \; \;  \;  \; \alpha_{i\!+\!2}  \!<\! \alpha_{i\!-\!1}\!-\!\alpha_i\!+\!\alpha_{i\!+\!1} \label{eq:ineq1} ~~~(iii)  \end{cases} \\
\alpha_i& \!>\!\alpha_{i\!-\!1}:\begin{cases}
                                      \alpha_{i}\!-\!\alpha_{i\!-\!1}  \!<\! \alpha_{i\!+\!1}\!+\!\alpha_{i\!+\!2} ~~~~~~~~~(i) \\
                                      \; \; \; \; \;  \; \;  \alpha_{i\!+\!2}  \!<\! \alpha_{i}\!-\!\alpha_{i\!-\!1}\!+\!\alpha_{i\!+\!1} ~~~(ii)\\
                                      \; \; \; \; \;  \; \; \alpha_{i\!+\!1}  \!<\! \alpha_{i\!+\!2}\!+\!\alpha_{i}\!-\!\alpha_{i\!-\!1}  ~~~(iii) \label{eq:ineq2} \end{cases}
\end{align}
As with the unique folds, these inequalities imply that generic vertices have two binding folds (denoted $b$ and $b'$) that straddle a common plate (denoted $B$).  Looking specifically at Eqs.~\ref{eq:ineq1}(ii) and \ref{eq:ineq2}(ii), we see that if fold $i$ is binding then either $i+1$ or $i-1$ is unique.  This is consistent with Fig.~\ref{fig:uniqueBinding}, where we see that, on a particular branch, the binding fold always has its corresponding unique fold next to it.

\begin{figure}[t!]
\includegraphics[]{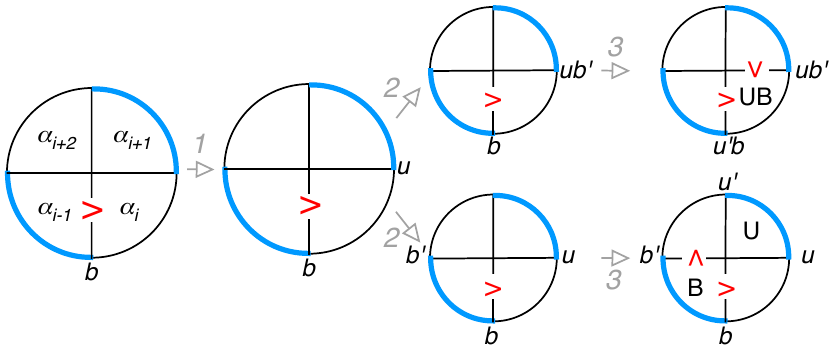}
\caption{(Color online) Relationships involving the dominant pair. The dominant pair is indicated by the thick blue rim.  We assume that fold $i$ is binding and that  $\alpha_{i-1}>\alpha_{i}$ (the case for $\alpha_{i-1}<\alpha_{i}$ is similar \cite{footnote3}). It follows from Eqs.~\ref{eq:ineq1} and \ref{eq:ineq2} that (1) fold $i+1$ is unique, and plates $i+1$ and $i-1$ are the dominant pair, and (2) there are two possibilities for the $b'$ folds.  Once the location of $b'$ is determined, (3) the locations of the $u'$ fold and the $U$ and $B$ plates are fixed according to Eqs.~\ref{eq:ineq1} and \ref{eq:ineq2}; this leads to two subtypes with either $U$ and $B$ the same (top) or opposite (bottom).}
 \label{fig:dominantPair}
\end{figure}

\paragraph{The dominant pair and generic subtypes:}
Equations \ref{eq:ineq1}(iii) and \ref{eq:ineq2}(iii) involve the two pairs of opposing plates, specifically addressing for which pair the sum of the sector angles is largest (and thus larger than $\pi$).  We call this pair the {\em dominant pair}, and we can use it to quickly identify the unique and binding plates of a generic vertex.  Figure \ref{fig:dominantPair} illustrates that the connection between unique and binding folds can be reformulated as follows: if fold $i$ is a binding fold then the neighboring fold separated by the smaller of the neighboring plates is the corresponding unique fold, $u$, and the inequalities expressed in Eqs.~\ref{eq:ineq1}(iii) and \ref{eq:ineq2}(iii) lead to a definite arrangement of the dominant pair.  By then considering the two possible locations for the other binding fold, $b'$, we can easily establish two different possible folded configurations for plate $U$, plate $B$, and the dominant pair (Fig.~\ref{fig:dominantPair}). It follows by inspection that the sector angle for plate $B$ is a local extremum. If the $B$ plate is a local minimum, the $U$ and $B$ plates are equal, and the $U$ plate is not part of the dominant pair.  If plate $B$ is a local maximum, the $U$ and $B$ plates are opposite and the $U$ plate is part of the dominant pair.

\begin{figure}[t!]
\includegraphics[]{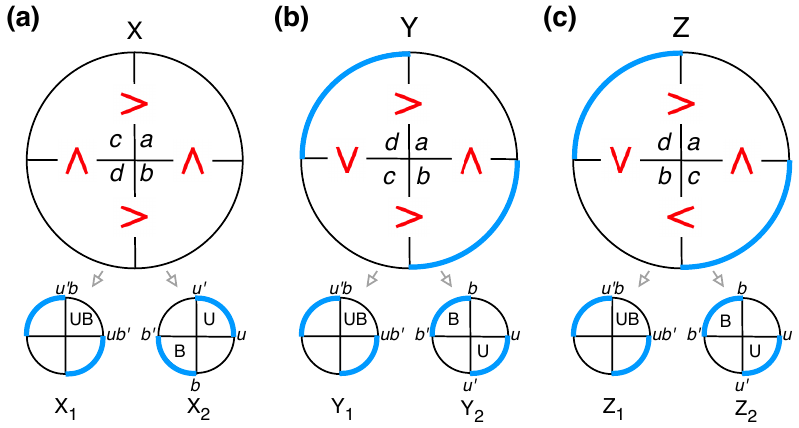}
\caption{(Color online)  Subtypes and folding branches for generic vertices.  For each type/subtype, the dominant pair is indicated by the thick blue rim and the unique/binding plates by $U/B$.  The folding branches on each subtype correspond either to folds $u$ being unique and $b$ binding, or $u'$ being unique and $b'$ binding.}
 \label{fig:genericSubtypes}
\end{figure}

In Fig.~\ref{fig:genericSubtypes}, we use the rules regarding the dominant pair to show that for each generic type there are two possible arrangements of the unique and binding folds on the first branch ($u$ and $b$) and on the second branch ($u'$ and $b'$), This reveals that generic vertices come in two subtypes depending on the relative locations of $U$ and $B$, which we will call subtype 1 when they are the same (here the $B$ plate is always a local minimum) and subtype 2 when they are opposite (here the $B$ plate is always a local maximum).

We note in passing that the subtype classification is closely related to the Grashof classification of (spherical) 4-bar linkages.  A Grashof linkage is one where (at least) one of the output bars (equivalent to plates) can rotate continuously, which is useful for performing engineering tasks.  This is possible if the sum of the shortest and longest bars is greater than the sum of the remaining two. This translates to $a+d<b+c$, and for all three generic vertices this is precisely the condition for subtype 1. Hence, even though spherical linkages and origami are different in that (a) the vertices here have sector angles that add to $2\pi$ and (b) cannot self intersect \cite{belcastro:2002}, subtype 1 vertices are related to (non-intersecting) Grashof linkages \cite{Chiang:1984, Barker:1985}.

\section{Special 4-vertices}

Special vertices arise when sector angles or sums of pairs of sector angles are equal.  Incorporating such constraints changes the combinatorics in determining the possible types of vertices, ultimately resulting in 16 new ones.  Furthermore, this allows for the possibility that some of the inequalities that govern the unique and binding folds become equalities, which changes the nature of the folding motion and, in some cases, the number of folding branches.  For example, the well-studied class of flat-foldable 4-vertices occurs when there is no dominant pair, i.e.,~when the sums of even and odd sector angles are equal. In that case, all folds close simultaneously. Clearly, qualitatively different rules apply in vertices such as these.  (In principle, one could consider more exotic constraints such as products, etc.  However, as can be verified by looking at the vertex folding equations, the motion depends solely on sums and differences.  This means that the relevant relationships correspond to individual sector angles or sums thereof being equal \cite{Huffman:1976, Evans:2015, Waitukaitis:2015}.)

In this section, we systematically determine all special vertex types.  We subdivide these based on the number of incorporated constraints, resulting in codimension-1, -2, and -3 special vertices.  We show how the relationships amongst these and the generic vertices can be visualized by looking at regions, lines and points in 2D subspaces of the full, 3D space of sector angles.  Finally, we analyze each special type and determine the locations of their unique and binding folds and further characterize their folding branches.

\begin{figure}[t!]
\includegraphics[]{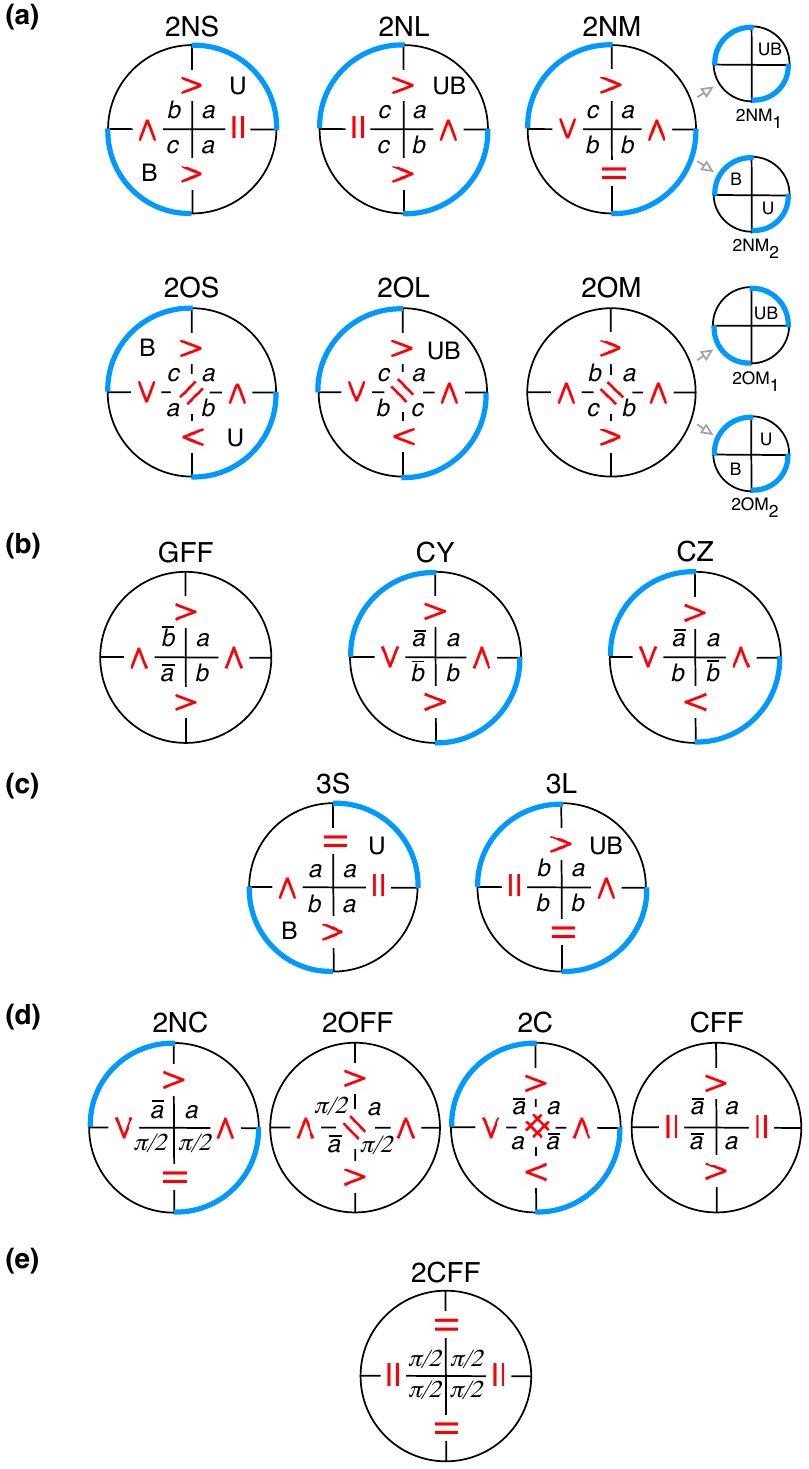}
\caption{Special 4-vertices.  Six codimension-1 special types arise if two sector angles are equal (a).  Three more arise when sums of pairs are equal (b).  Codimension-2 vertices include ones with three sector angles equal (c), and also those with two equal and sums of pairs equal (d).  The only codimension-3 vertex has all sector angles equal to $\pi/2$ (e).  The relative sizes are indicated both by letters ($a<b<c<d$) and also by equal and less than/greater than signs.  For vertices where sums of pairs are equal, the complementary pair is denoted by a barred angle, e.g., $\bar{a}+a=\pi$.  In (a) and (c), the unique and binding plates are well-defined ($U/B$), and in all but the types with FF the dominant pair (thick blue rim) is well-defined.  For all vertices except 2NM and 2OM, the subtype is predetermined by the constraints.}
\label{fig:specialTypes}
\end{figure}

\begin{figure*}[t!]
\includegraphics[width=\textwidth]{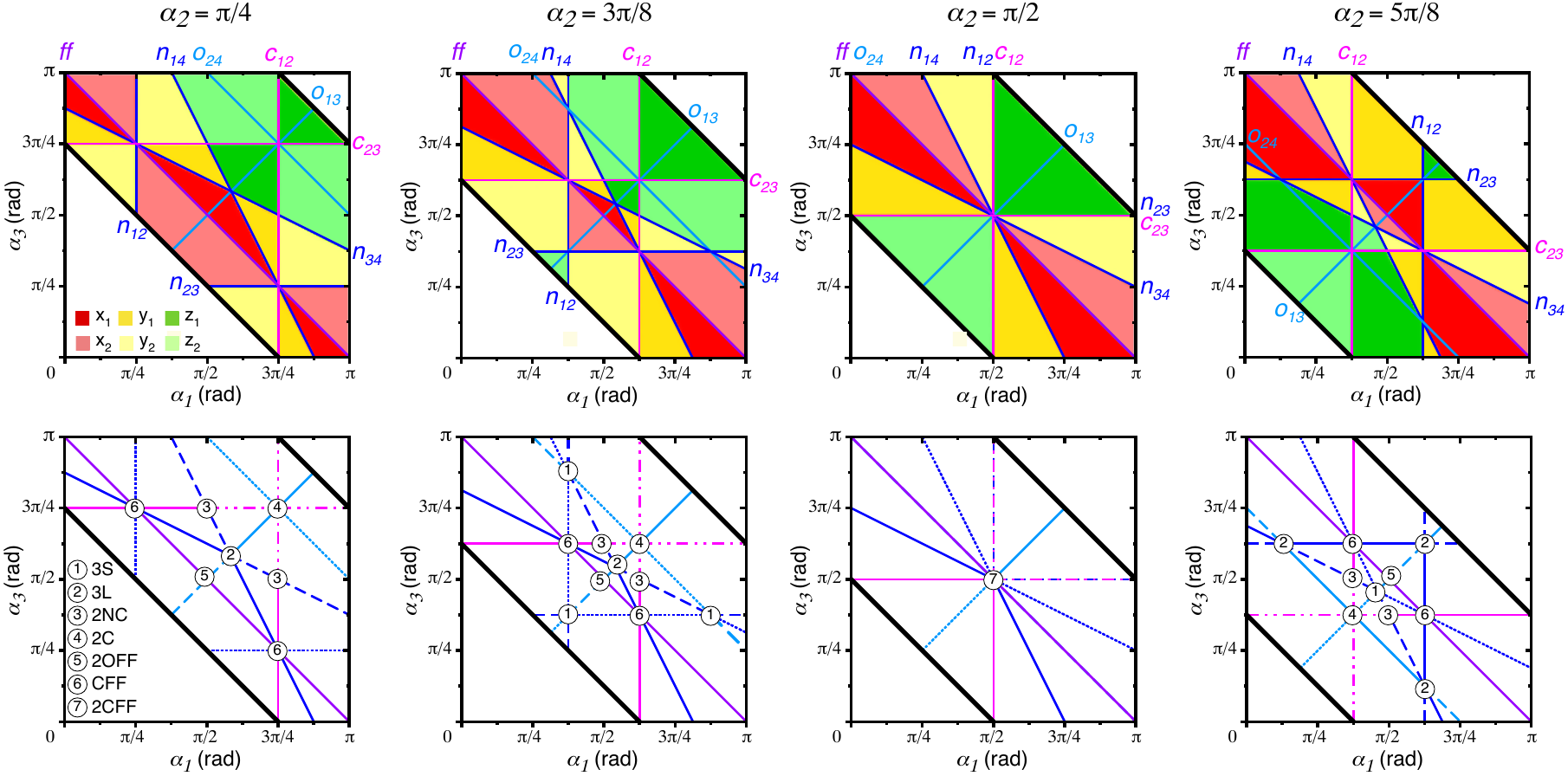}
\caption{Visualizing the full space of 4-vertices.  The top row shows planes of constant $\alpha_2$ (values indicated above).  White areas are forbidden as one sector angle becomes $<0$ or $>\pi$.  The colors denote the generic type (red=X, yellow=Y, green=Z) and subtypes (darker=1, lighter=2).  Nine lines partition the generic types and subtypes: $n_{ij}$ where neighboring sector angles $\alpha_i=\alpha_j$ (four dark blue lines); $o_{ij}$ where opposite sector angles $\alpha_i=\alpha_j$ (2 light blue lines); $c_{ij}$, or collinear lines, where the sum of the neighboring sector angles $\alpha_i+\alpha_j=\pi$ (2 pink lines); and {\em ff} where $\alpha_1+\alpha_3=\alpha_2+\alpha_4=\pi$ (purple line).  The bottom row recasts this based on the codimension-1, -2 and -3 special vertex types.  Lines (codimension-1) are colored as before, but here dots correspond to smallest angles equal, dashes to middle angles equal, and solid lines to largest angles equal.  The flat-foldable line is unchanged, but special types CY and CZ now have solid and dash-dot-dot lines, respectively.  The line intersections corresponding to codimension-2 and -3 vertices are indicated by the legend in the leftmost panel. For an animation showing the full 3D space of slices, see the supplemental video \cite{supplVideo}.}
\label{fig:typeSlice}
\end{figure*}

\paragraph{Codimension-1 vertices:}
The simplest of the codimension-1 special vertices occurs when two sector angles are equal.  This disrupts the well-defined ordering of generic vertices in one of three ways:  the equal angles can be the smallest ($a=b$), the middle ($b=c$) or the largest ($c=d$).  For each of these possibilities, the equal angles can be arranged next to each other or opposite to each other.  This makes it clear that six special vertex types are associated with two angles being equal, as in Fig.~\ref{fig:specialTypes}(a).  We give these the names 2NS, 2NL, 2NM, 2OS, 2OL, and 2OM, where the ``2'' stands for two angles being equal, ``N/O'' distinguishes whether or not the equal angles are arranged {\underline n}ext to or {\underline o}pposite each other, and ``S/L/M'' identifies them as the {\underline s}mallest, {\underline l}argest or {\underline m}iddle.  Like generic vertices, our notions of the $U$ plate, the $B$ plate, and the dominant pair remain well-defined for these vertices.  Unlike generic vertices, however, only 2NM and 2OM vertices have two subtypes, whereas for all other special vertices the subtype is fixed via the constraints.

Setting the sums of pairs of angles equal results in more codimension-1 special types.  Like generic vertices, these still have a well-defined ordering of the sector angles $a<b<c<d$.  The equal sums must therefore be $a+d=b+c=\pi$. (In the language of spherical linkages these would be classified as {\it change point mechanisms} \cite{Barker:1985}).  The preserved ordering signifies that these come in three varieties that are derived directly from each of the generic types.  One of these is the familiar flat-foldable vertices, which we denote GFF as they are the most {\underline g}eneral ones possible.  Interestingly, GFF vertices are only derived from generic type X.  The other two types have the angles that sum to $\pi$ arranged next to each other, which results in two opposing folds lying on precisely the same line.  We give these the names CY and CZ, with ``C'' denoting that they have {\underline c}ollinear folds and ``Y/Z'' signifying the parent generic type.  As with most of the special types generated from setting two angles equal, the additional constraints here preclude distinct subtypes.  We will show in the next section that these vertices require modifications to our notions of the $U$ and $B$ plates.

\paragraph{Codimension-2 and codimension-3 Vertices:}
Adding another constraint results in codimension-2 vertices, and the simplest of these occurs when three sector angles are equal.  We can quickly determine that there are just two of these by inserting two equalities into our ordered angles, i.e.,~$a=b=c<d$ or $a<b=c=d$.  We designate these 3S and 3L.  As with vertices where two angles are equal, these still have a well-defined dominant pair and $U$ and $B$ plates.

Codimension-2 vertices can also arise via the combination of two equal angles and equal sums of pairs of angles; this results in four types.  In two of these, the angles that are equal also add to $\pi$---hence they are $\pi/2$.  They can be arranged next to each other, which results in another collinear vertex (2NC), or they can be arranged opposite from each other, which results in a subclass of flat-foldable vertices (2OFF).  Alternatively, if the equal angles do not add to $\pi$, then the result is a double collinear vertex, 2C, or a collinear vertex that is also flat-foldable, CFF.  This last type, CFF, is the base vertex for the Miura-ori.  (In the midst of the wide variety of other vertices we have identified, it should be clear that this type occupies a very small region of phase space.)  Adding any additional constraints results in the singular codimension-3 vertex, 2CFF, where all sector angles are equal to $\pi/2$.

\begin{table}[]
\begin{tabular}{|l|l|l|l|l|}  \hline
$n_{ij}$ &  $o_{ij}$ & $c_{ij}$ & ~{\em ff}~ & Type\\ \hline
&   &   &   & X$_1$, X$_2$, Y$_1$, Y$_2$, Z$_1$, Z$_2$ \\ \hline
$\bullet$ &   &   &   & 2NS, 2NM, 2NL, 3S, 3L\\
& $\bullet$ &   &   & 2OS, 2OM, 2OL \\
&   &$ \bullet$ &   & CY, CZ \\
&   &   & $\bullet$ & GFF\\ \hline
$\bullet$ &   & $\bullet$ &   & 2NC \\
& $\bullet$ &   & $\bullet$ & 2OFF \\
& $\bullet$ & $\bullet$ &   & 2C \\
$\bullet$ &   & $\bullet$ & $\bullet$ &CFF \\ \hline
$\bullet$ & $\bullet$ & $\bullet$ & $\bullet$ & 2CFF \\
\hline
\end{tabular}
\label{t9}
\caption{Connections among special vertices.  Each vertical block indicates a different vertex codimension, with the vertex types as indicated.  The vertical columns indicate which conditions are met: $n_{ij}$ for equal neighboring angles, $o_{ij}$ for equal opposite angles, $c_{ij}$ for collinear folds and {\em ff} for a flat-foldability.}
\end{table}

\paragraph{Visualizing the full geometric space:}  The schematics in Figs.~\ref{fig:genericTypes} and \ref{fig:specialTypes} are useful for identifying vertex types, but to see the connections between different types we now more carefully consider the configuration space spanned by the $\alpha_i$.  As the sum of sector angles equals $2\pi$, generic Euclidean vertices occupy the bulk of a 3D space.  We can specify a point in this space using $\alpha_{1,2,3}$ as coordinates ($\alpha_4\!:=\!2\pi\!-\!\sum_{i=1,3}\alpha_i$).  The special vertices reside on nine distinct planes (codimension-1), six lines (codimension-2) and one point (codimension-3).

While we cannot easily depict this 3D space, we gain insight by looking at 2D slices corresponding to fixing one sector angle, as presented in Fig.~\ref{fig:typeSlice}. (For a video traversing the full 3D space, see the supplemental material \cite{supplVideo}.)  We see the layout of generic vertices by coloring each region according to type and shading by subtype.  The codimension-1 special vertices, which occupy planes in the full space but lines in this 2D space, are now seen to create divisions between generic types and subtypes.  For example, the boundaries between generic types are delineated by four lines $n_{ij}$, corresponding to adjacent folds $i$ and $j$ being equal.  The different generic subtypes are bounded by the lines {\em ff}, $c_{12}$ and $c_{23}$, which correspond to the flat-foldable vertices ($\alpha_1+\alpha_3=\alpha_2+\alpha_4$), and the two possible arrangements for collinear folds ($\alpha_i+\alpha_j=\pi$ for $c_{ij}$).

In the lower panels, we connect this to the 16 special types as previously identified.  The lines $n_{ij}$ and $o_{ij}$ are split into the special types where two angles are equal, while the lines {\em ff}, and $c_{ij}$ are split into the special types corresponding to equal sums.  The intersection points of lines corresponding to codimension-2 and codimension-3 vertices bring to light particularly interesting features.  For example, 3S and 3L vertices live at the intersections of different $n_{ij}$ and $o_{ij}$ lines, and any of the three generic types can be reached via an infinitesimal deviation away from these highly constrained domains.  As another example, the plots make it readily apparent that 2C vertices live only in generic type Z.  These relationships and more are summarized in Table I, which highlights several other interdependencies: e.g., that flat-foldable vertices ({\em ff}) that have two equal neighboring plates ($n_{ij})$ are automatically collinear ($c_{ij}$).

\subsection{Special folding behavior}

\begin{figure}[b!]
\includegraphics[]{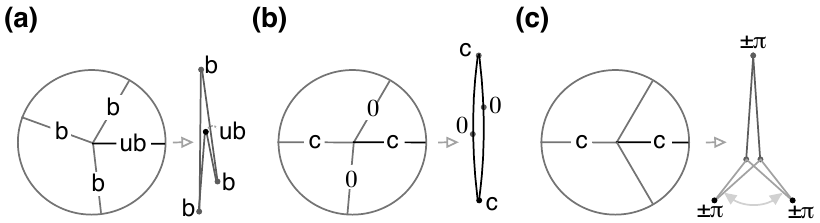}
\caption{Special folding motions.  (a)  2D sketch showing that in flat-foldable vertices, all four folds are binding and unique folds remain well-defined.  (b)  For collinear vertices, one branch of folding motion corresponds to the two collinear folds (denoted by $c$) being equal while the remaining null folds (denoted by $0$) stay flat.  Our planar polygon representation fails for this folding motion and we therefore add slight curvature to the plates.  (c)  For a vertex with collinearity and flat-foldability, additional folding branches emerge when the collinear folds bind at $\pm\pi$ and the previously null folds fall on top of each other and permit further motion.}
\label{fig:specialFolding}
\end{figure}

\begin{figure}[]
\includegraphics[]{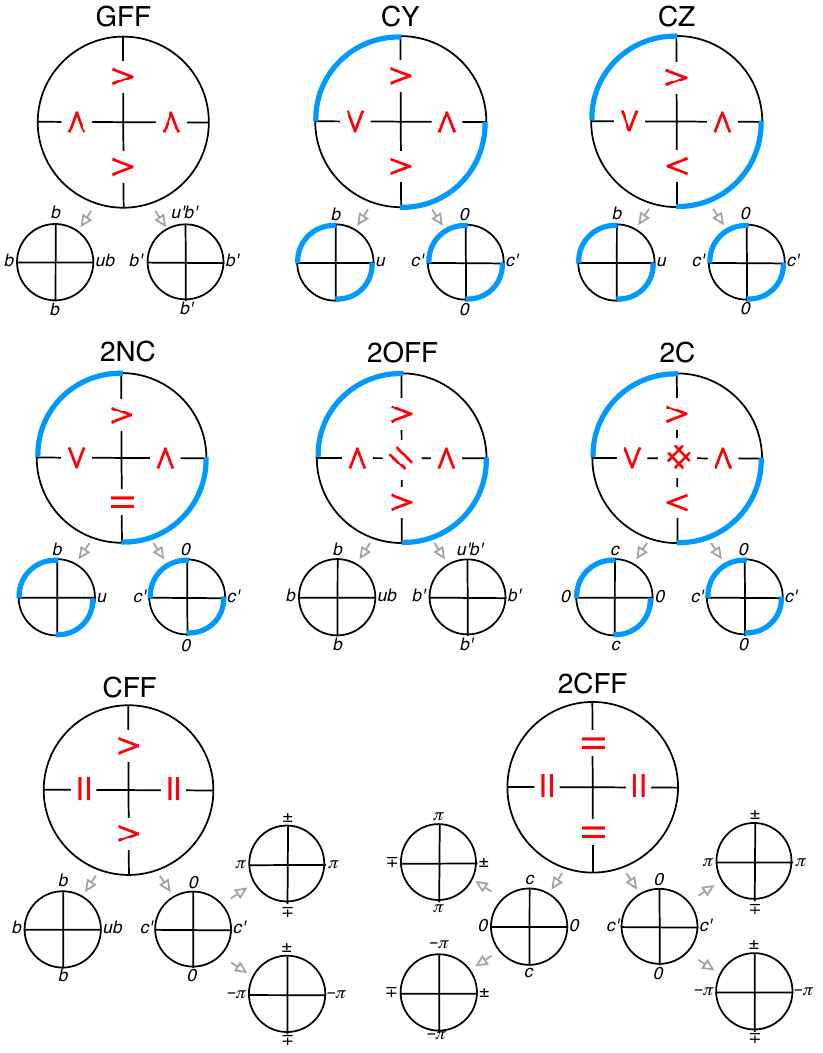}
\caption{Vertices with special folding branches.  For vertices that are flat-foldable, all folds are capable of binding on all branches---this applies to types GFF, 2OFF, CFF, and 2CFF.  For vertices that have collinearity, one branch corresponds to the collinear folds (denoted with a $c$) folding simultaneously while the null folds (denoted with a $0$) remain flat. Types CY, CZ, 2NC, and CFF have a single collinearity, while types 2C and 2CFF have two.  Finally, vertices that have reflection symmetry across collinear folds develop new branches when the collinear folds reach $\pm\pi$.  For these branches, the two previously null folds align and can vary continuously between $\pm\pi$, but have opposite signs, here indicated by $\pm$ and $\mp$.  We encourage the reader to use the paper cutouts in the supplemental material to illustrate these concepts \cite{supplCutouts}.}
\label{fig:specialBranches}
\end{figure}

We now analyze the folding behavior for special vertices.  For types where two angles are equal [Fig.~\ref{fig:specialTypes}(a)] or three angles are equal  [Fig.~\ref{fig:specialTypes}(c)], the dominant pair and $U/B$ plates are well-defined.  This implies the folding behavior of the two branches can be deduced in the same way as for generic vertices.  All of the other special types, however, involve either flat-foldability, collinearity, or a combination thereof, and this creates substantive changes in their folding behavior.  We now consider the specific modifications that arise in each of these cases.  In Fig.~\ref{fig:specialBranches}, we take these modifications into account and sketch out all folding branches for each of the affected special types.

\paragraph{Flat-foldable vertices:}  Flat-foldable vertices are, by definition, those where all folds are binding folds. This becomes apparent  in the maximally folded state, illustrated for example in
Fig.~\ref{fig:specialFolding}(a), where we draw a nearly closed GFF vertex.  Flat-foldability alone does not change what we have previously said regarding the unique plate---this is because the angles that add to $\pi$ are opposite from each other.  For generic flat-foldable (GFF) vertices, as well as for 2OFF vertices, there is a well-defined unique plate and two branches of folding motion, but all folds are capable of binding  --- see Fig.~\ref{fig:specialBranches}.

\paragraph{Collinear vertices:}  Collinear vertices have two opposite folds aligned, and this modifies the behavior of one of the folding branches.  As illustrated in Fig.~\ref{fig:specialFolding}(b), collinear vertices have one branch where two of the folds are fixed flat, i.e., they have a value of zero throughout the entire folding motion.  We call this the {\em collinear branch}, and we distinguish here the {\em collinear folds}, which close together, from the {\em null folds} which remain flat. Note that both of the collinear folds will bind during this folding motion.  The other branch behaves normally, and thus we conclude that a single collinearity introduces two null folds and two binding folds --- see types CY,  CZ and 2NC in Fig.~\ref{fig:specialBranches}. A double collinearity (as in 2C) simply leads to two collinear branches --- each of the folds can be binding, and there are no unique folds.

\paragraph{Collinear and flat-foldability:}  Finally, consider what happens when there is both flat-foldability and collinearity.  This situation results in CFF or 2CFF vertices with reflection symmetry across the pair of collinear folds (see table I), as in Fig.~\ref{fig:specialFolding}(c).  Once a collinear branch is maximally folded, the previously null folds align on top of each other, and this enables new branches of motion.  Here the collinear folds remain fixed at either $\pm \pi$, while the previously null folds are free to vary from $-\pi$ to $\pi$.  During this motion these fold angles have equal magnitude but opposite sign; thus in some sense they are unique, but only on these new branches.  For type CFF, there is reflection symmetry across a single collinear fold, thus two new branches are introduced, giving rise to four branches in total.  For type 2CFF, there is reflection symmetry across both pairs of collinear folds, which means these have six branches of folding motion---see Fig.~\ref{fig:specialBranches}.

\section{Summary and Outlook}

Motivated by the connection between geometry and mechanical functionality, we have fully characterized the folding motion of rigid Euclidean 4-vertices.  First, we have shown that there are three generic vertex types, which correspond to the three unique ordered arrangements of the sector angles around the vertex center.  Generic vertices have two unique folds that straddle a common plate $U$ and two corresponding binding folds that straddle a common plate $B$.  This implies that generic vertices have two branches of motion.  By drawing connections between the unique folds, binding folds, and dominant pair, we have also shown that generic vertices come in two subtypes depending on whether the $U$ and $B$ plates are the same or opposite to each other.  

By introducing constraints between the sector angles, we have shown that there are 16 different types of special vertices.  The simplest of these are the codimension-1 vertices, which occur either when two angles are equal (resulting in six special types) or when sums of pairs of angles are equal (resulting in three special types).  Adding one more constraint results in codimension-2 vertices, which occur when three angles are equal (two types) or when two angles are equal and sums of pairs are equal (four types).  Finally, there is just one codimension-3 vertex corresponding to all sector angles equal to $\pi/2$.  For vertices that have flat-foldability, collinearity, or reflection symmetry across folds, the folding motion is dramatically changed---the notions of the binding fold and unique folds must be modified and new branches of folding motion can emerge.  In principle, similar organizational schemes could be used to categorize and study vertices with more folds, or which are non-Euclidean.

Finally, we would like to stress that  the majority of recent work on origami metamaterials has focussed on tilings of the codimension-2 CFF vertex, which exhibit very interesting and useful behaviors.
Our results highlight the wide range of generic and non generic 4-vertices, and we suggest that much can be gained by looking at materials made from these other types of vertices.

{\em Acknowledgements}  We thank C.~Coulais, R.~Menaut, P.~Dieleman, C.~Santangelo, and A.~Evans for productive discussions and support from NWO via a VICI grant. This work is part of the research programme of the Foundation for Fundamental Research on Matter (FOM), which is part of the Netherlands Organisation for Scientific Research (NWO).

\bibliographystyle{apsrev}

\begin{thebibliography}{36}
\bibitem{footnote1}  If one sector angle is equal to the other three, it is equal to $\pi$. In this case the vertex has one simple branch of motion where the two aligned folds that straddle this sector angle simply open or close simultaneously.  If one angle is larger than the other three, the vertex is fixed.
\bibitem{footnote2}  The distinction between non-primed and primed folds is arbitrary, but will become useful when we determine the corresponding binding folds.
\end{thebibliography}

\end{document}